\documentclass[notoc,paper,letterpaper]{JHEP3}
\usepackage{amssymb}
\def\be{\begin{equation}}
\def\ee{\end{equation}}
\def\bea{\begin{eqnarray}}
\def\eea{\end{eqnarray}}
\def\ie{{\it i.e.}}

\def\chk{$^\surd$}

\def\U{\mbox{U}}

\def\SO{\mathop{\rm SO}}

\def\Sp{\mathop{\rm Sp}}
\def\bo{\hbox{1\kern-.23em{\rm l}}}

\def\til{\widetilde}

\def\del{{\partial}}
\def\a{{\alpha}}
\def\b{{\beta}}
\def\t{{\tau}}
\def\x{{\xi}}
\def\o{{\omega}}

\title{Classification of N=2 Superconformal Field Theories 
with Two-Dimensional Coulomb Branches, II}

\author{Philip C. Argyres and John R. Wittig\\ 
Physics Dept., U. Cincinnati, Cincinnati OH 45221-0011\\
\email{argyres,jwittig@physics.uc.edu}}

\abstract{We continue the classification of 2-dimensional 
scale-invariant rigid special Kahler (RSK) geometries.  
This classification was begun in \cite{acsw0504} 
where singularities corresponding to curves of the form 
$y^2=x^6$ with a fixed canonical basis of holomorphic one forms 
were analyzed.  Here we perform the analysis for the $y^2=x^5$ 
type singularities.  (The final maximal singularity type, 
$y^2=x^3(x-1)^3$, will be analyzed in a later paper.)  
These singularities potentially describe the Coulomb 
branches of $N{=}2$ supersymmetric field theories in four 
dimensions.  We show that there are only 13 solutions 
satisfying the integrability condition (enforcing the RSK 
geometry of the Coulomb branch) and the $Z$-consistency 
condition (requiring massless charged states at singularities).  
Of these solutions, one has a marginal deformation, 
and corresponds to the known solution for certain $\Sp(2)$ 
gauge theories, while the rest correspond to isolated strongly 
interacting conformal field theories.}
 
\begin{document}

\section{Introduction}

The classification of all possible $N=2$ superconformal
field theories (SCFTs) with one complex dimensional Coulomb
branches (``rank one" theories) was carried out almost a decade ago 
\cite{sw94,apsw9511,mn96}.  It corresponds to a classification
of all one-dimensional scale-invariant rigid special 
Kahler (RSK) geometries, and coincides with the Kodaira 
classification \cite{kodaira} of complex singularities of 
families of elliptic curves.

The classification of all possible rank 2 $N=2$ SCFTs turns out 
to be considerably more complicated, but still 
amenable to a systematic computation.  In this paper we continue 
this  classification, begun in \cite{acsw0504}.  We are interested 
in classifying possible interacting SCFTs which are neither IR 
free nor factor into lower-rank interacting and/or IR free parts.  
These correspond to the ``maximal" singularities of the 
corresponding genus 2 Seiberg-Witten curve.  As discussed in 
\cite{acsw0504}, there are just three types of such maximal 
singularities corresponding to curves of the form $y^2=x^6$, 
$y^2=x^5$, and $y^2=x^3(x-1)^3$ with a fixed canonical basis 
of holomorphic one-forms.  The first case was analyzed in 
\cite{acsw0504}, we analyze the second case here, and 
we leave the third case to another paper.

As reviewed in \cite{acsw0504} the effective action on 
a two-dimensional Coulomb branch is described by a
genus 2 Riemann surface together with basis of
holomorphic 1-forms.  All genus 2 Riemann surfaces are
hyperelliptic, and can be described algebraically
as a double-sheeted cover of the complex $x$-plane
(plus infinity) by $y^2=P(x)$ where $P$ is a polynomial
of order 5 or 6 in $x$.  We can choose the complex 
coordinates $x$ and $y$ such 
that the basis of holomorphic one-forms has the
canonical form $\o_u=xdx/y$ and $\o_v=dx/y$.

The central charge, $Z$, of the $N=2$ superalgebra depends 
linearly on the magnetic and electric charges of the 
$\U(1)^2$ low energy gauge group on the Coulomb branch of
the moduli space.  It is related to the holomorphic one-forms 
by
\be\label{h1form}
\del_u Z = \oint x dx/y ,\qquad
\del_v Z = \oint dx/y.
\ee
Here $u$ and $v$ are global complex coordinates on the 
Coulomb branch.  The integrability of (\ref{h1form}) gives
the partial differential equation for the curve,
\be\label{inteqn}
\del_u y^{-1}-\del_v (x y^{-1}) = \del_x (b y^{-1}), 
\ee
where $b$ is an arbitrary meromorphic function of $x$.
(As we discuss in section 3 below, one can show that
$b$ is in fact a quadratic polynomial in $x$.)  At genus
2 this integrability equation completely encodes the
constraints on the Coulomb branch geometry coming from
$N=2$ supersymmetry (\ie, its RSK geometry).

To search for SCFTs, we look for scaling solutions of the 
integrability condition (\ref{inteqn}).  Scaling means
that $u$, $v$, $x$ and $y$ can be assigned scaling
dimensions (which we denote by square brackets).  For the
scaling to make sense, all scaling dimensions must
have positive real parts.  It is convenient to define
$r := [v]/[u]$ and $s := 1/[v]$. 
By (\ref{h1form}) and (\ref{inteqn}) the scaling dimensions 
of all quantities can be expressed in terms of $[u]$ and 
$[v]$, and therefore $r$ and $s$.
We then define dimensionless variables, 
$\x := u^{1-r} x$, 
$\o := -r u^{-r} v$,
$\eta := u^{1+rs-2r} y$, and
$\b := r^{-1} u^{2-r} b +(1-r^{-1})\xi$,
in terms of which the integrability equation becomes
\be\label{inteqn2}
[(\x-\o)\del_\o+(s-1)]\eta^{-1} = \del_\x(\b \eta^{-1}).
\ee

The form of $\eta$ and $\b$ can be greatly constrained
by using the remaining freedom to make changes of
variables among $x$, $u$, and $v$, as well as by
imposing that the solutions are single-valued in
$u$ and $v$.  These conditions are outlined in section
2 for the $y^2=x^5$ type singularities.  The result is that they 
restrict $r$ to take a discretely infinite set of values for
each possible value of $s$, and they allow the differential
equation (\ref{inteqn2}) to be reduced to a set of
polynomial equations.

We solve these polynomial equations for $\eta$ (and $\b$) 
in section 3, finding 28 solutions whose form depends only on
$s$.  The list of solutions is given in table 1.  Note
that each entry corresponds to an infinite number of solutions,
since $r$ can take infinitely many values.

The modulus of the central charge gives the lower bound on the mass 
for any state with corresponding electric and magnetic charges.
Singularities in the effective action are the result of 
charged states becoming massless, therefore every singularity 
must be accompanied by vanishing central charge.  This 
``$Z$-consistency condition" is an extra physical requirement 
on our solutions.  In the scaling case when the integrability 
condition (\ref{inteqn2}) is satisfied, (\ref{h1form}) can be 
integrated to give
\be\label{Zeqn}
Z = \frac{u^{rs}}{rs} \oint \frac{(\x-\o)}{\eta} d\x .
\ee
The $Z$-consistency condition is then evaluated in section 4
by evaluating (\ref{Zeqn}) at the various singularities of the
$\eta$'s found from solving the integrability equation.  The
result is that only 13 solutions survive; they are recorded in table 
\ref{tab.cft} below.

We conclude in section 5 with a discussion of these scale-invariant
solutions for the possible $N=2$ supersymmetric low energy 
effective action on 2-dimensional Coulomb branches.  All are 
consistent with $N=2$ superconformal invariance.  One has an 
exactly marginal operator, and coincides with the known curve 
\cite{as9509} for certain $\Sp(2)$ scale invariant $N=2$ supersymmetric 
gauge theories.  Of the remaining 12 solutions, only one has been 
previously identified \cite{ad9505,ehiy9603,eh9607} as an $N=2$ 
superconformal fixed point theory found by appropriately tuning 
vevs, masses, and couplings in other $N=2$ field theories.  The 
other 11 are new isolated fixed point theories.

\section{$\bf y^2=x^5$ type singularities}

The degenerations of the general genus 2 curve $y^2=P(x)$, 
where $P$ is an order six polynomial in $x$, can be classified
according to how the six branch points collide with one another.
These correspond to all the ways of partitioning the six branch
points into colliding subsets.  Of these, there are just three
maximal degenerations, which have the property that every cycle
on the Riemann surface is homologous to a vanishing cycle.  They
are the partitions (6), (5,1), and (3,3).  The first corresponds
to degenerations where all six branch points collide at a single
point in the projective $x$-plane; the second to 5 branch points 
colliding at a single $x=x_1$ while the sixth remains separate
at $x=x_2$; and the third to three colliding at $x_1$ and the 
other three colliding at $x_2$.  As mentioned above, this paper
is devoted to the second, or (5,1), case.

Even after fixing the basis of holomorphic one-forms, there
is left unfixed a group of reparametrizations of $x$, $u$, and
$v$, called the ``holomorphic reparameterizations" in 
\cite{acsw0504}, which act by general holomorphic reparameterizations 
on $u$ and $v$, together with a fractional linear transformation on 
$x$.  This can be used to send 
any three distinct points on the $x$ projective plane to chosen 
values.  We can partially fix this freedom by choosing the
maximal (5,1) degeneration to be at $u=v=0$, and by choosing
$x_1=0$ and $x_2=\infty$.  This leaves a singularity 
of the form $y^2 \sim x^5$ at $u=v=0$.  Therefore, we can write 
the curve as
\be
y^2 = a(u,v)\left( f_6(u,v)x^6 + x^5 + \sum_{k=0}^4 f_k(u,v)x^k \right),
\ee
for some unknown coefficients $a$ and $f_k$, where the $f_k$
vanish when $u=v=0$.

Single-valuedness of the physics as a function of the good
coordinates $(u,v)$ on the moduli space imply \cite{acsw0504}
that the $f_k$ are single-valued functions on the Coulomb
branch; $a$, however, need not be single-valued.

The scaling hypothesis together with our other coordinate
choices mentioned above leaves only a three-parameter subgroup
of the holomorphic reparameterizations unfixed.  One of these
is simply an overall rescaling, which can be used to set the
overall coefficient of $a$ to 1.  Another
is the freedom to shift $v\to v+Cu^r$ with an associated shift
of $x\to x-rCu^{r-1}$.  This freedom can be completely
fixed by setting $f_4(u,v)=0$ by an argument analogous to one
used in \cite{acsw0504}.  The remaining unfixed holomorphic 
reparametrizations involves a rescaling of $v$ and $x$ keeping
$u$ unchanged; it can be fixed by setting any non-zero coefficient
in one of the $f_k$'s to 1.

A happy simplification in for the (5,1) type singularities is
that scaling, regularity, and the vanishing of the $f_k$
at $u=v=0$ implies that $f_6$ must vanish identically.
Thus, in the dimensionless scaling variables introduced
in section 1, the (5,1) curve can be taken to be of the
form
\be\label{etaeq}
\eta^2 := \a(\o) \phi(\x,\o) \quad\mbox{with}\quad
\phi(\x,\o) = \xi^5 + \sum_{k=0}^3 \phi_k(\o) \xi^k .
\ee
Scaling plus regularity on the moduli space imply that
the functions $\phi_k(\o)$ have the following dependences
on $\o$:
\bea\label{listphi}
\phi_3 &=& a_3 \o + b_3 \nonumber\\
\phi_2 &=& a_2 \o^2 + b_2 \o + c_2 \nonumber\\
\phi_1 &=& a_1 \o^3 + b_1 \o^2 + c_1 \o + d_1 \nonumber\\
\phi_0 &=& a_0 \o^4 + b_0 \o^3 + c_0 \o^2 + d_0 \o + e_0 .
\eea 

A further simplification compared to the $y^2=x^6$ case
is in the argument determining the $x$-dependence in
the unknown integration function $b(u,v,x)$ that appears
in the integrability condition (\ref{inteqn}).  In
\cite{acsw0504} the highest order in $x$ of $y^2$ was six;
in the present case it is only five.  Since the highest 
order of $b$ in $x$ was shown to be no more than the highest 
order of $y^2$ divided by two, and since $b$ is single-valued
in $x$, we find that $b$ is at most quadratic in $x$: 
$b = \sum_{k=0}^2 b_k(u,v)\, x^k$, or, in terms
of the dimensionless scaling variables, $\b = \sum_{k=0}^2
\b_k(\o)\,\xi^k$. 

\section{Solutions to the integrability equation}

Substituting the first equation in (\ref{etaeq}) into 
(\ref{inteqn2}) gives the scale invariant integrability equation 
\be
(\x-\o)\del_\o\phi = \b \del_\x\phi - 
\phi[2(1-s)+2\del_\x\b + (\x-\o)\del_\o \ln \a ].
\ee
Substituting the $\xi$ expansions of $\phi$ and $\b$
then give a series of ordinary differential equations
in $\o$ for the coefficient functions $\a$, $\phi_k$ and $\b_k$.
In this case we find that the first non-identically zero 
coefficient of $\x$ gives
\be\label{alphaeq}
\del_\o \ln \a  = \b_2 ,
\ee
allowing us to eliminate $\a$. 
Two of the resulting equations allow us to simply solve
for $\b_0$ and $\b_1$ in terms of $s$, $\b_2$, and $\phi_3$:
\be\label{betaeq}
5\b_0 = 2\b_2\phi_3 +\phi_3' ,\qquad
3\b_1 = 2(1-s) -\o\b_2 .
\ee
Upon substituting these equations into the integrability 
equations we are left with the following 4 equations:
\bea
0 &=& \b_2  [\phantom{-75\phi_0 +{}} 25\o\phi_0  
+\phantom{1}6\phi_1\phi_3 ] + 
[-50(1-s)\phi_0 \phantom{{}-15\phi_0'}+15\o\phi_0' +3\phi_1\phi_3' ]
\nonumber\\
0 &=& \b_2  [-75\phi_0 +20\o\phi_1 +12\phi_2\phi_3 ] + 
[-40(1-s)\phi_1 -15\phi_0' +15\o\phi_1' +6\phi_2\phi_3' ]
\nonumber\\
0 &=& \b_2  [-60\phi_1  +15\o\phi_2  +18\phi_3\phi_3 ] + 
[-30(1-s)\phi_2 -15\phi_1' +15\o\phi_2'  +9\phi_3\phi_3' ]
\nonumber\\
0 &=& \b_2  [-45\phi_2  +10\o\phi_3 \phantom{{}+18\phi_3\phi_3}] + 
[-20(1-s)\phi_3 -15\phi_2'  +15\o\phi_3'\phantom{{}+9\phi_3\phi_3'} ].
\eea
Finally, eliminating $\b_2$, substituting (\ref{listphi}), and 
expanding in powers of $\o$ gives a set of polynomial equations 
for the coefficients $\{a_i,b_i,c_i,d_i,e_0\}$ and $s$.  

Solving this system of polynomial equations and using (\ref{alphaeq}) 
and (\ref{betaeq}) then determines $\a$, $\b$, $\eta^2$, and $s$.
It turns out that this polynomial system, unlike the one found for
the $y^2=x^6$ type singularity in \cite{acsw0504}, can be solved
completely in a reasonable amount of time on a desktop computer.
The method we employed was simply judicious elimination of variables
using (at worst) resultants of pairs of polynomials.  This
resulted in a large (about $10^3$ node) tree of possibilities.
(The interested reader can request a copy of a 
{\sl Mathematica}{\footnotesize\texttrademark}
notebook with the computation from the authors.)

\TABLE[ht]
{\begin{tabular}{||l||l|l|l||} \hline
\#		&$[v]$	&$\phi$		             &$\a$      
\\ \hline\hline
1\chk	&1		&$\x^5 + \t_1 \x^3 + \t_2 \x^2 + \t_3 \x + 1$ &1
\\ \hline
2      &8/7	&$128(9\x-10)^4(9\x-5) + 1440(9\x-10)^3\o$  &$\o^{-2}$
\\
		&		&$\ \mbox{}+1620(9\x-10)^2\o^2+729(9\x-10)\o^3$ &
\\ \hline
3		&5/4	&$\x^5 + \o$                           &$\o^{-1/5}$
\\ \hline
4		&5/4	&$625(9\x-10)^4(9\x-5)+7500(9\x-10)^3\o$  &$\o^{-2}$
\\ 
		&		&$\ \mbox{}+1620(9\x-10)^2\o^2+7290(9\x-10)\o^3+2187\o^4$ &
\\ \hline
5		&4/3	&$\x(\x^4+\o)$ & $\o^{-1/4}$
\\ \hline
6		&4/3	&$16(9\x-10)^4(9\x-5)+30(9\x-10)^2(18\x-35)\o$  &$\o^{-1}$
\\
		&		&$\ \mbox{}-225(54\x-55)\o^2$ &
\\ \hline
7		&4/3	&$8(9\x-10)^4(9\x-5)+60(9\x-10)^2(18\x-17)\o$   &$\o^{-1}$
\\ 
		&		&$\ \mbox{}+45(27\x-29)\o^2$ &
\\ \hline
8\chk	&10/7	&$\x^5+5\x-4\o$	                    &$1$
\\ \hline
9\chk	&8/5	&$\x(\x^4+4\x-3\o)$	                 &$1$
\\ \hline
10		&5/3	&$\x^5+\o^2$	                        &$\o^{-2/5}$
\\ \hline
11		&40/21	&$\x^5+\o(5\x-4\o)$	                 &$\o^{-1/4}$
\\ \hline
12		&2		&$\x(\x^4+\o^2)$	                    &$\o^{-1/2}$
\\ \hline
13\chk	&20/9	&$\x^5 + 10\x^2\o + 15\x\o^2 + 6\o^2$ &$\o^{-2/5}$
\\ \hline
14\chk	&12/5	&$\x(\x^4+\o(4\x-3\o))$		          &$\o^{-1/3}$
\\
15\chk	&12/5	&$(\x^2+2\o)(\x^3+3\x\o+2\o)$         &$\o^{-1/3}$
\\ \hline
16		&5/2	&$\x^5+\o^3$		                    &$\o^{-3/5}$
\\ \hline
17\chk	&5/2	&$\x^5+(5\x-3\o)^2$	                 &1
\\ \hline
18		&20/7	&$\x^5+\o^2(5\x-4\o)$	             &$\o^{-1/2}$
\\ \hline
19\chk	&15/4	&$\x^5+\o(5\x-3\o)^2$	             &$\o^{-1/3}$
\\ \hline
20		&4		&$\x(\x^4+\o^3)$	                    &$\o^{-3/4}$
\\ \hline
21\chk	&4		&$\x[\x^4+\t\x^2(\x-\o/2)+(\x-\o/2)^2]$ &1
\\ \hline
22\chk	&24/5	&$\x(\x^4+\o^2(4\x-3\o))$	         &$\o^{-2/3}$ 
\\ \hline
23		&5		&$\x^5+\o^4$	                       &$\o^{-4/5}$
\\ \hline
24		&40/7	&$\x^5+\o^3(5\x-4\o)$	            &$\o^{-3/4}$ 
\\ \hline
25\chk	&15/2	&$\x^5+\o^2(5\x-3\o)^2$	            &$\o^{-2/3}$ 
\\ \hline
26\chk	&8		&$\x(\x^4+\o(2\x-\o)^2)$	            &$\o^{-1/2}$ 
\\ \hline
27\chk	&10		&$\x^5+(5\x-2\o)^3$	               &1 
\\ \hline
28\chk	&20		&$\x^5+\o(5\x-2\o)^3$	            &$\o^{-1/2}$ 
\\ \hline
\end{tabular}
\caption{Potentially physical solutions  the integrability equation
for curves $\eta^2=\a\phi$ of $y^2=x^5$ type singularity.  A check 
next to the row number means the corresponding solution passes the 
$Z$-consistency condition at $v=0$. \label{tab.crv}}}

Table \ref{tab.crv} lists the solutions, $\eta^2=\a\phi$, of the 
integrability equation together with their associated value of $[v]$.
In addition to the curves in this table, there are also formal 
solutions found with $[v]$ = $-8$, $-5$, 4/5, 1, 4/3, 3/2, 2, 3, 4, 
$\infty$, and one with arbitrary $[v]$.  The ones with $[v]$ = 4/5, 
1, 4/3, 3/2, 2, 3, 4, as well as the curve with arbitrary $[v]$, 
were discarded because they do not resolve the singularity on the 
Coulomb branch (\ie, they remained singular for all $u$ and $v$);  
Seiberg-Witten curves have a well-defined physical interpretation 
describing an $N=2$ $\U(1)^2$ low energy effective action only when 
they are non-singular.   The solutions with $[v]$ = -8, -5, and 
$\infty$ were discarded simply because they do not have a consistent 
scaling interpretation.

\section{Z-consistency condition}

It is necessary to test the physicality of remaining solutions by 
applying the condition that every singularity is accompanied by
vanishing $Z$ (for some choice of $\U(1)^2$ electric and magnetic
charges).  This condition follows from the fact that $|Z|$ 
gives a lower limit on the mass of charged states and that singularities 
occur when charged states become massless; for more detail see 
\cite{sw94}.  These singularities occur in three ways.

\paragraph{Singularities along $v = 0$.}

Only the checked curves in table \ref{tab.crv} pass the Z-consistency
test along the submanifold $v=0$ emanating from the singularity at
the origin.  They pass it either by simply having no singularity
at $v=0$ when $u\neq 0$, or because $Z$ indeed vanishes there for
some choice of integration contour.  The curves which fail this check
categorized into two groups.  The first group (curves numbered 3, 5,
10, 11, 12, 16, 18, 20, 23, and 24) have the form $\eta^2 = \o^{-j/(5-k)}
[\x^5 + c\o^j\x^k + \ldots]$ where $\x^k$ is the highest power of
$\x$ appearing in the curve after $\x^5$.  In all these curves, the
singularity at $\o=0$ occurs at $\x=0$ on the curve.
The central charge can then be approximated around $\o = 0$ by 
\be
Z \sim \o^{-j/(5-k)}\oint dx(\o^{j/(5-k)}x-\o) (x^5+\ldots)^{-1/2},
\ee
after making the change of variables $x=\o^{j/(k-5)} \x$.  The
leading $\o$-dependence then exactly cancels, leaving $Z$ finite
as $\o\to0$.

The second group (curves numbered 2, 4, 6, and 7) fails 
because of the property that at $\o = 0$, the singularity occurs at 
$\x \neq 0$.  This removes the positive contribution from the 
numerator of the integrand, which prevents the central charge from 
vanishing.  This is identical to what happens in the next section 
when $\x_0 \neq \o_0$ (as it should since we could always use our 
holomorphic reparameterization freedom to shift $\o$ and $\x$ to 
move the singularity away from $\o = 0$).

\paragraph{Singularities along $v \sim u^r$.}

Along with singularities at $\o = 0$, it is possible for 
singularities to occur at finite $\o = \o_0$.  As $\o \to \o_0$ 
branch points of $\phi$ may collide at $\x = \x_0$ so that 
$\phi \sim (\x-\x_0)^{2+n}\til\phi$ where $\til\phi$ is nonsingular
and $n$ is a some non-negative integer.
The value of $Z$ can be analyzed around $(\x_0,\o_0)$ by making 
the change of variables  $\x = \x_0 + y^{1/2}x$,  $\o = \o_0 + y$, 
and taking $y \to 0$.  Upon making this transformation,
\be
Z \sim y^{-n/4}
\oint dx (\x_0-\o_0 +y^{1/2}x-y ) \til\phi^{-1/2},
\ee
which generically remains finite or even diverges as $y \to 0$.  
However if $\x_0 = \o_0$ an extra factor of 
$y^{1/2}$ comes from the numerator of the integrand and
$Z \sim y^{(2-n)/4}$.  In order for the exponent 
to be positive we must have $n=0$ or 1.  Examination of the checked
solutions in table \ref{tab.crv} shows that all have the
properties $\x_0 = \o_0$ and $n=0$, and therefore all pass this
check.

\paragraph{Singularities along $u = 0$.}

Finally, we must check the $Z$-consistency condition at any
singularities along the $u=0$ submanifold, which corresponds to
$\o=\infty$ in dimensionless variables.  All but two of the
checked curves in table \ref{tab.crv} have the form $\eta^2 = 
\o^{-j/(5-k)} [\x^5+c\o^j\x^k+\ldots+c'\o^\ell\x^m]$ where $\x^k$
is the highest power of $\x$ occurring (after $\x^5$) and $\x^m$
is the lowest.  Changing variables to $x = \o^{\ell/(m-5)}\x$, gives
at large $\o$
\be\label{4.3}
Z = \frac{u^{rs}}{rs}\o^{{j\over 2(5-k)}+1-{3\ell\over 2(5-m)}}
\oint dx (x^5+\ldots)^{-1/2} .
\ee
Recall, that at fixed $v$, $\o \sim u^{-r}$.  Therefore, the 
exponent of $u$ in the previous expression is $s - {j\over2(5-k)} 
- 1 + {3\ell\over2(5-m)}$.  Remarkably, for all of these curves 
this exponent vanishes, implying that $Z$ remains finite as $u\to0$.  
Thus all these curves fail the $Z$-consistency test at $u=0$ unless 
they happen to have no singularity along the $u=0$ submanifold.  
Whether the curves are singular or not at $u=0$ is controlled by 
whether the term $\o^\ell \x^m$ with the lowest power of $\x$ 
vanishes there or not.  Reintroducing dimensionful quantities, 
this term reads $u^{5r-5}\o^\ell\x^m \sim u^{r(5-\ell-m)-(5-m)}
v^\ell x^m$, so $r$ must assume the value
\be
r = \frac{5-m}{5-\ell-m}
\ee
for the curve not to be singular at $u=0$.  This condition selects 
a single value of $r$ for each curve.  But existence of a scaling 
limit implies also that $r<1$ \cite{acsw0504}.  This is satisfied
for this value of $r$ for all the curves except curve number 1 (with
$[v]=1$).  

The two checked curves which do not follow the above pattern are
numbers 13 and 15.  (Their prefactor $\a$ is not proportional to
$\o^{-j/(5-k)}$.)  Nevertheless, the same argument works for these
curves:  $Z$ is non-zero at $u=0$, and only for a single value of
$r$ are they non-singular there.

The resulting list of the 13 solutions which pass the $Z$-consistency
test is given below in table \ref{tab.cft}.  Here we have recorded
the values of $[v]$ and $[u]$, put back in the explicit $u$ and $v$ 
dependence in the curve, and used the holomorphic reparametrization
rescaling and shift freedom to simplify the form of the curves 
where possible.

\TABLE[ht]
{\begin{tabular}{||rr|rl||} \hline
\multicolumn{2}{||l|}{Dimensions}&\multicolumn{2}{l||}{Curve}\\
$[v]$&$[u]$&\multicolumn{2}{l||}{$y^2= \ldots$}\\ \hline\hline
10/7 &8/7  &           &$[x^5+(ux+v)]$\\ \hline 
8/5  &6/5  &           &$[x^5+x(ux+v)]$\\ \hline 
20/9 &10/9 &$v^{-2/5}$ &$[x^5+v(5ux^2-15vx-6vu)]$\\ \hline
12/5 &6/5  &$v^{-1/3}$ &$[x^5+vx(2ux+3v)]$\\ \hline
12/5 &6/5  &$v^{-1/3}$ &$[x^2-4v][x^3-2v(3x+2u)]$\\ \hline
5/2  &3/2  &           &$[x^5+(ux+v)^2]$	\\ \hline
15/4 &3/2  &$v^{-1/3}$ &$[x^5+v(2ux+3v)^2]$\\ \hline
4    &2    &           &$[x^5+\t x^3(ux+v)+x(ux+v)^2]$\\ \hline
24/5 &6/5  &$v^{-2/3}$ &$[x^5+v^2x(ux+3v)]$\\ \hline  
15/2 &3/2  &$v^{-2/3}$ &$[x^5+v^2(ux+3v)^2]$\\ \hline
8    &2    &$v^{-1/2}$ &$[x^5+vx(ux+2v)^2]$\\ \hline 
10   &4    &           &$[x^5+(ux+v)^3]$\\ \hline
20   &4    &$v^{-1/2}$ &$[x^5+v(ux+2v)^3]$\\ \hline
\end{tabular}
\caption{Solutions of the integrability equation and
$Z$-consistency condition for $y^2=x^5$ singularities. 
For all curves the basis of holomorphic one-forms is 
$\o_u=xdx/y$ and $\o_v=dx/y$.\label{tab.cft}}}

\section{Discussion}

All of these curves have $[u]$ and $[v]$ real and greater
than one, consistent with an interpretation as interacting
$N=2$ superconformal fixed points.  Only one, the
curve with $[v]=4$ and $[u]=2$ has a dimensionless
coupling $\t$.  This curve in fact corresponds to the
known curve \cite{as9509} of the $\Sp(2) \simeq \SO(5)$ 
scale invariant gauge theory with either six massless
hypermultiplets in the $\bf 4$ or three in the $\bf 5$
of the gauge group.  (Both theories' effective actions
are described by the same curve in the massless limit, an
occurrence which is known to occur in other theories
\cite{sw94}.)   The weak coupling limit occurs as
$\t\to\pm2$, in which case $v$ and $u$ can be identified
with the adjoint casimirs for $\Sp(2)$, explaining their
dimensions.

One other curve, the one with $[v]=10/7$ and $[u]=8/7$
has been found previously \cite{ad9505,ehiy9603,eh9607}
by tuning parameters in higher-rank asymptotically free
$N=2$ theories.  Presumably many other curves in table
\ref{tab.cft} can also be found in this way, but a systematic
search along these lines is algebraically prohibitively 
complicated.

\section*{Acknowledgments}
It is a pleasure to thank M. Crescimanno and A. Shapere for useful 
discussions and comments.  We are also grateful for the hospitality
of the School of Natural Sciences at the Institute for Advanced Study, 
where some of this work was carried out.
PCA and JW are supported in part by DOE grant DOE-FG02-84ER-40153.  
PCA was also supported by an IBM Einstein Endowed Fellowship, and  
JRW by the Hanna Fellowship from U. Cincinnati.

\end{document}